\documentclass[aps,prb,twocolumn,superscriptaddress,floatfix]{revtex4-1}
\usepackage{hyperref}
\usepackage{graphicx}
\usepackage{amsfonts}
\usepackage{amssymb}
\usepackage{amsmath}
\usepackage{mathrsfs}
\usepackage{txfonts}
\usepackage{lipsum}
\usepackage{color}
\usepackage{wasysym}
\usepackage{letltxmacro}

\DeclareMathOperator{\Real}{Re}
\DeclareMathOperator{\Imag}{Im}

\DeclareMathOperator{\Diag}{diag}

\newcommand{\refE}[1]{Eq.~(\ref{#1})}
\newcommand{\rem}[1]{} 

\begin{document}
\title{Two-level system as topological actuator for nanomechanical modes}

\author{C. Dutreix}
\affiliation{Universit\'e de Bordeaux and CNRS, LOMA, UMR 5798, F-33400 Talence, France}

\author{R. Avriller}
\affiliation{Universit\'e de Bordeaux and CNRS, LOMA, UMR 5798, F-33400 Talence, France}

\author{B. Lounis}
\affiliation{Institut d'Optique $\&$ CNRS, LP2N UMR 5298, F-33400 Talence, France}
\affiliation{Universit\'e de Bordeaux, LP2N, F-33400 Talence, France}

\author{F. Pistolesi}
\affiliation{Universit\'e de Bordeaux and CNRS, LOMA, UMR 5798, F-33400 Talence, France}

\begin{abstract}
We investigate theoretically the dynamics of two quasidegenerate mechanical modes coupled through an open quantum two-level system.
A mean-field approach shows that by engineering the retarded response of the two-level system with a coherent drive, the non-Hermitian mechanical spectrum exhibits an exceptional degeneracy point where the two modes coalesce.
We show that this degeneracy can be exploited to manipulate the vectorial polarization of the mechanical oscillations.
We find that adiabatically varying the detuning and the intensity of the drive induces a rotation of the mechanical polarization, which enables the topological and chiral actuation of one mode from the other.
This topological manifestation of the degeneracy is further supported by quantum-jump Monte Carlo simulations to account for the strong quantum fluctuations due to the spontaneous emission of the two-level system. 
Our presentation focuses on a promising realization based on flexural modes of a carbon-nanotube cantilever 
coupled to a single-molecule electric dipole irradiated by a laser. 
\end{abstract}

\maketitle

\section{Introduction}
The manipulation and detection of nanometer oscillators are important challenges in nanomechanics \cite{clerk2010introduction,poot2012mechanical,aspelmeyer2014cavity}, and recent progress has led to unprecedented high-resolution sensors
\cite{chan2001quantum,rugar2004single,lassagne2008ultrasensitive,chaste2012nanomechanical,moser2013ultrasensitive,abbott2016observation}.
Most nano-oscillators exhibit a multimode dynamics
\cite{conley2008nonlinear, eichler2012strong, zhang2012synchronization, antoni2013nonlinear, faust2013coherent, shkarin2014optically}.
In particular, the flexural dynamics of suspended nanowires involves nearly degenerate orthogonal modes, which enables the detection of anisotropic and nonconservative force fields \cite{gloppe2014bidimensional,cadeddu2016time,de2017universal}. 
Such advances in vectorial force microscopy rely on vectorial oscillations whose control is crucial to scan surfaces \cite{rossi2017vectorial}.
Strategies to accurately monitor ultralight cantilevers, such as carbon nanotubes that allow $zN/\sqrt{\text{Hz}}$ force sensitivity \cite{de2018ultrasensitive}, are then highly desirable to develop more sensitive vectorial probes.
One interesting possibility that we investigate here is to exploit the nonconservative force induced by the detection system.
Indeed, open systems can exhibit intriguing degeneracy points in the analytic continuation of their spectra, associated with the coalescence of the eigenstates and known as exceptional points (EPs) \cite{mondragon1996berry,heiss1999phases,berry2004physics,holler2018non}.
EPs have recently allowed efficient topological energy transfers between two harmonic modes of a membrane placed in the middle of an optical cavity \cite{xu2016topological}.
Two-level systems (TLSs) constitute minimal quantum systems to detect and manipulate mechanical motions \cite{o2010quantum,PhysRevLett.121.183601}. 
For example, their coupling to flexural modes allowes the localization of emitters randomly 
distributed in micropillars \cite{yeo2014strain,de2017strain}. 
It was also shown that single-molecule TLSs are sensitive local probes to measure, through the Stark effect, the small 
displacements of a charged nanotube \cite{puller2013single,pistolesi2018bistability}.
Using a TLS to detect and actuate nanomechanical oscillators brings two fundamental differences 
with respect to the use of optical or electromagnetic cavities.
({\em i}) A strongly pumped optical cavity has a linear behavior, whereas the quantum nature of a TLS is intrinsically nonlinear.
({\em ii}) In highly populated optical cavities the Poissonian fluctuations are negligible, whereas the TLS experiences strong quantum fluctuations due to spontaneous emission.
Whether it is possible to observe EPs in the electromechanical spectrum of a cantilever coupled to a TLS naturally appears as a fundamental question.
In this paper, we show that adiabatically by varying along a closed path the frequency and the intensity of a coherent field driving a TLS coupled to two quasi-degenerate mechanical modes it is possible to induce a change in the state of the mechanical oscillator that depends on the topology of the path.
We show that this behavior is due to the presence of an EP in the mean-field description of the electromechanical spectrum.
Using quantum-jump Monte Carlo simulations, we prove that this property holds in the presence of strong TLS fluctuations.
The  presence of the EP allows one to generate elliptic mechanical eigenmodes, whose axis angles can be controlled. 
Finally, we propose a detection scheme to probe the topological switch between the two quasidegenerate flexural modes in single-molecule spectroscopy. 
%

\section{Mechanical modes coupled to a driven TLS}

\subsection{The system Hamiltonian}
We consider the generic Hamiltonian of a TLS driven by a coherent field, linearly coupled to two nearly degenerate mechanical modes,
\begin{align}\label{Hamiltonian0}
\mathcal{H} =
& -\frac{\omega_{\rm TLS}}{2} \hat\sigma_{z} + \Omega_{\rm L} \cos(\omega_{\rm L}t)\hat\sigma_{x} + 
\omega_{1} b^{\dagger}_{1}b_{1} + \omega_{2} b^{\dagger}_{2}b_{2} \notag \\
& - \sum_{i=1,2}  
g_{i} (b_{i}^{\dagger}+b_{i}) \hat\sigma_{z}
\,.
\end{align}
The operators $\sigma_{x}$ and $\sigma_{z}$ are Pauli matrices and describe the
TLS of energy splitting $\omega_{\rm TLS}$ ($\hbar=1$).
The TLS is driven by a coherent field of frequency $\omega_{\rm L}$ and intensity (Rabi frequency) $\Omega_{\rm L}$.
It couples with strength $g_i$, the mechanical modes
of frequencies $\omega_i$, and destruction (creation) operators $b_{i}$ ($b^\dagger_{i}$)\cite{puller2013single}.
The Hamiltonian $\mathcal{H}$ describes the unitary evolution.
We further consider dissipation processes.
The driven TLS can spontaneously emit photons toward the electromagnetic environment with decay rate $\Gamma$.
The mechanical modes are coupled to thermal baths and have damping rates $\gamma_i$ \cite{pistolesi2018bistability}.
Both baths are assumed to have the same temperature $T_{0}$.

\subsection{Electronanomechanical system}
\begin{figure}[t]
    \centering
    \includegraphics[trim = 280mm 180mm 290mm 5mm, clip, width=6cm]{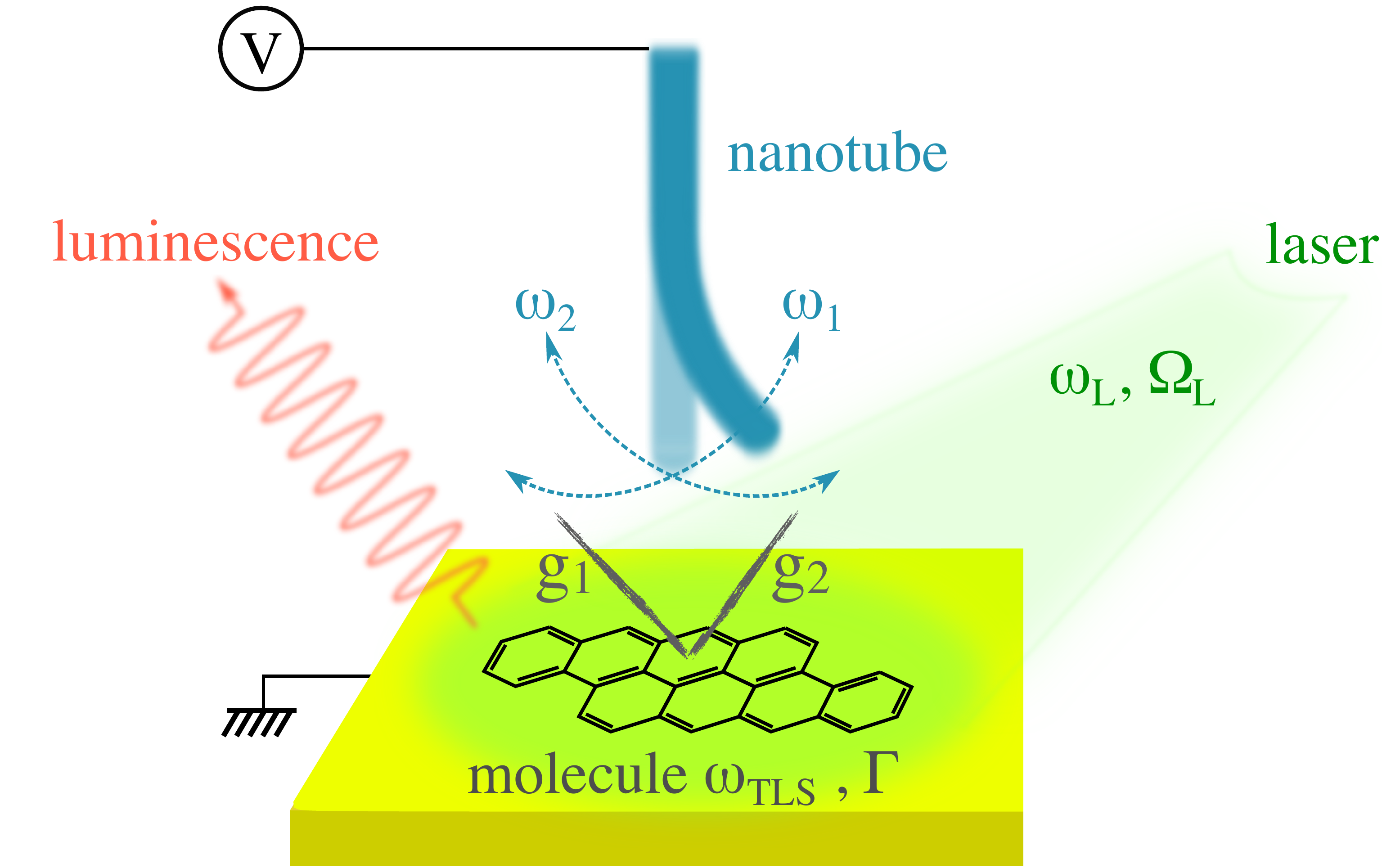}
    \caption{\textbf{Electronanomechanical system:}
    two orthogonal fundamental modes of a suspended carbon nanotude couple electrically to the TLS of a single molecule pumped by a far-field laser. 		See, also, Ref.~\onlinecite{puller2013single} for more details. }
    \label{fig:system}
\end{figure}

The generic model introduced above can describe various physical systems \cite{yeo2014strain,Ania2014,de2017strain}.
We will focus our presentation on the system proposed in Ref.\onlinecite{puller2013single}  and shown schematically in 
Fig.\,\ref{fig:system}.
%
%
In this case, the TLS is given by the electronic doublet in organic molecules embedded in a solid-state matrix \cite{lounis1997single,brunel1998rabi}. 
As discussed in Ref. \onlinecite{puller2013single}, when a suspended carbon nanotube kept at a fixed difference of potential from
the substrate oscillates, it modulates the electric field on the electronic doublet, and by Stark effect it modulates the two-level system energy splitting. 
The coupling is proportional to the permanent electric dipole moment of the molecule, which can reach values up to two debyes \cite{Moradi2019Matrix}.
Performing molecular spectroscopy, then, allows one to detect the displacement of the oscillator.
In this paper we consider the presence of two  flexural modes that are quasidegenerate for symmetry reasons. 

Concerning the typical parameters one has that the carbon nanotube mass is $m\simeq10^{-20}$\,kg, 
 the fundamental frequencies satisfy $\omega_{i}/(2\pi)\simeq1$--$10$\,MHz with 
 quality factors $Q_i=\gamma_i/\omega_i\simeq10^{3}$--$10^{5}$ in the underdamped regime \cite{tsioutsios2017real}.
In usual single-molecule experiments performed at liquid-helium temperatures, the TLS exhibits a lifetime limited dephasing rate $\Gamma/2$, with $\Gamma/(2\pi)\simeq8$--$10\,$MHz.
Realistic values of the coupling strengths $g_i$ can be as large as $g_{c}\simeq 1 $GHz, which corresponds to a discharged electric field of about 10 mV/nm between the nanotube tip and the molecule substrate \cite{Smith_2005}.
%

\section{Mean-field mechanical dynamics}

\subsection{Langevin equation of motion}
Now we aim to describe the non-Hermitian dynamics of the mechanical modes
that occurs when the TLS and the environment are traced out. 
In usual cavity optomechanics, the coupling between the mechanical oscillator 
and the cavity field can be linearized to solve the problem exactly. 
This is not possible with a TLS due to its intrinsic nonlinear nature. 
On the other side, in our case, we can exploit the timescale separation between 
the mechanical and the TLS dynamics, $\gamma_i \ll\Gamma$.
We then follows  Refs.~\onlinecite{clerk2004quantum,clerk2010introduction} 
and derive a Langevin equation for the mechanical degree of freedom 
tracing out the TLS quantum degree of freedom.
This approach captures the Gaussian contributions and, in this sense, it is not limited to weak 
coupling. 
We obtain that the  expectation value of the displacement 
$ x_i= x^{\rm zpf}_{i}  \langle b_{i}+b^{\dagger}_{i} \rangle$, where $x^{\rm zpf}_{i}$
denotes the zero-point fluctuations, satisfies the Langevin equation,
\begin{eqnarray}
\lefteqn{
\ddot{x}_{i}(t) + \gamma_{i} \dot x_{i}(t) + \omega_{i}^{2} x_{i}(t) 
= 
\frac{g_{i}}{m x^{\rm zpf}_{i} } 
\langle \hat \sigma_{z} \rangle_{0} 
+ \frac{\delta F_{i}(t)}{m} 
}
\nonumber\\
&&
+ \frac{1}{m}  \sum_{j=1,2} \frac{g_{i} g_{j}}{x^{\rm zpf}_{i}x^{\rm zpf}_{j}  } \int \!\!dt' S_{\rm R}(t-t') x_{j}(t') \,.
\label{ClassicalEQM}
\end{eqnarray}
With $\langle \dots \rangle_{0}$, we indicate quantum averages evaluated in the absence of mechanical 
coupling ($g_{i}=0$).
The average force associated with $\langle \hat \sigma_{z} \rangle_{0}$ only shifts the equilibrium position of the mechanical oscillator and we disregard it from now on.
The forces $\delta F_i$ denote the Brownian thermal fluctuations, as well as the nonequilibrium stochastic fluctuations due to the spontaneous emission.
The last term describes the TLS-mediated retarded coupling between the mechanical modes.
It involves the retarded response function of the TLS in the presence of the laser field and the electromagnetic environment:
$
S_{\rm R}(t) = 
-i\theta(t)\langle [ \delta\hat\sigma_{z}(t), \delta\hat\sigma_{z}(0) ] \rangle_{0}
$, 
where $\delta\hat\sigma_{z}(t)=\hat\sigma_{z}(t)-\langle \hat\sigma_{z}(t)\rangle_{0}$ characterizes the fluctuations of the population difference.

\subsection{Non-Hermitian mean-field dynamics}
From \refE{ClassicalEQM},  we derive an effective non-Hermitian Hamiltonian that describes the oscillator's mode dynamics. 
We begin by neglecting the fluctuation forces $\delta F_i$.
We then linearize the equation of motion (\ref{ClassicalEQM}) for frequencies close to the 
two mechanical resonances $\pm \omega_i$ (see Appendix\,\ref{Derivation of the non-Hermitian Hamiltonian}).
We introduce the positive-frequency complex amplitude  ${\bf X}(t)$ from which one can obtain
the physical oscillator displacements, ${\bf x}=(x_1, x_2)=2 {\rm Re}[{\bf X}]$.
The ${\bf X}$ quantities formally obey a Schr\"{o}dinger equation
$i \dot {\bf X}=H {\bf X}$, 
where the effective Hamiltonian is non-Hermitian,
\begin{align}\label{HamiltonianEP}
H = 
\left(
\begin{array}{cc}
\omega_1-i\gamma_1/2-g_{1}^2 S_{\rm R}   &  -g_{1} g_2 S_{\rm R} \\
-g_{1} g_2 S_{\rm R}   &  \omega_2-i \gamma_2/2-g_{2}^2 S_{\rm R}
\end{array}
\right) .
\end{align}
Here, we assume that the mechanical frequency splitting is much smaller than the TLS linewidth, $\omega_{1}-\omega_{2}\ll\Gamma$.
Thus, the coupling between the nearly degenerate mechanical modes depends on the retarded response 
at the mean mechanical frequency $\omega_{0}=(\omega_{1}+\omega_{2})/2$, and so $S_{\rm R}\equiv S_{\rm R}(\omega_{0})$.

\begin{figure}[t]
    \centering
    \includegraphics[trim = 360mm 235mm 355mm 10mm, clip, width=6cm]{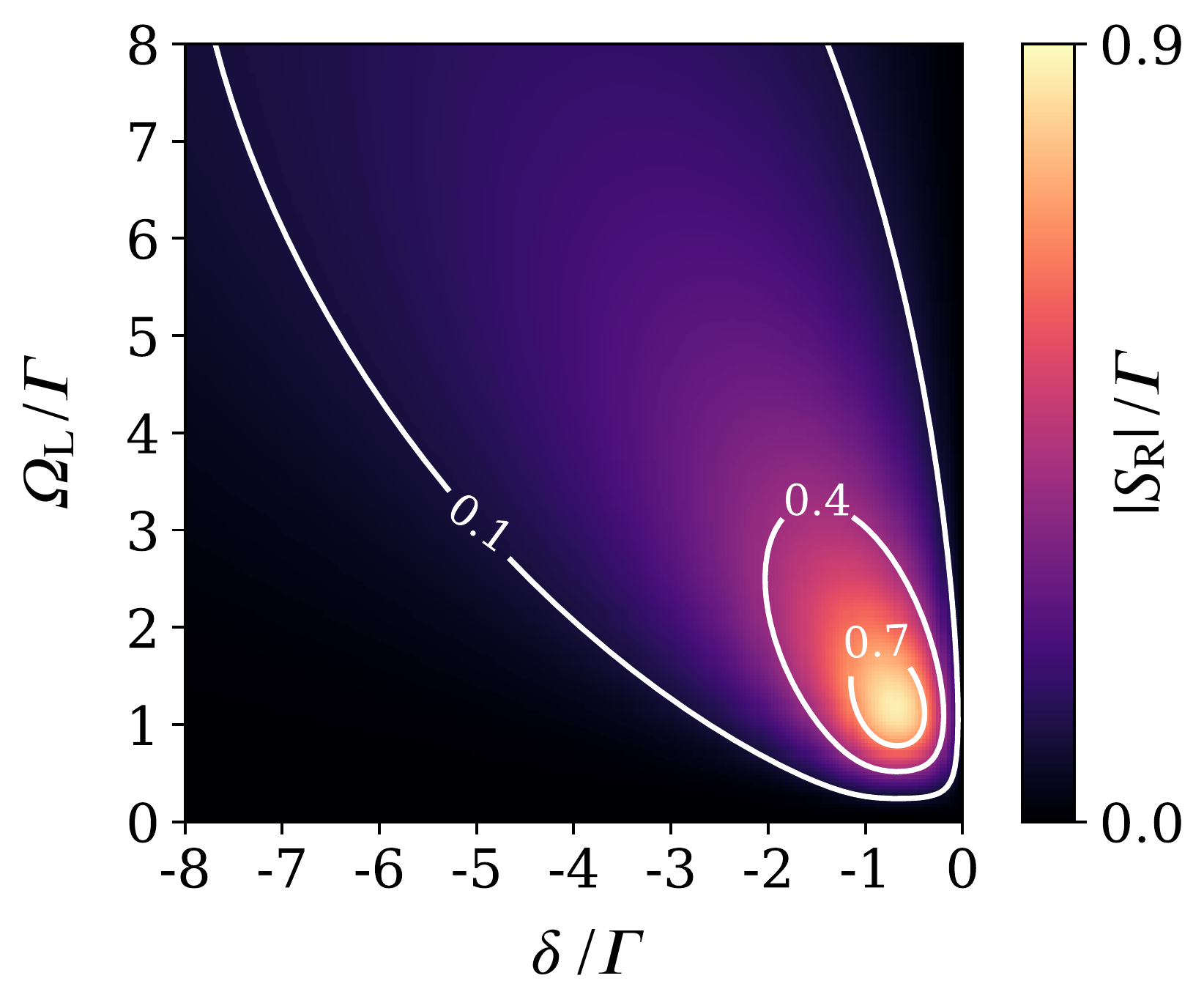}
    \caption{\textbf{TLS-mediated coupling}: strength of the retarded TLS response $|S_{\rm R}|$ as a function of the detuning $\delta$ and Rabi frequency $\Omega_{\rm L}$.}
    \label{PowerSpectrum}
\end{figure}

To determine the effective coupling between the mechanical modes, 
we need to evaluate $S_{\rm R}$.
This can be done by using a standard Born-Markov approximation for the 
evolution of the TLS reduced density matrix ${\rho}$ obtained by tracing out the electromagnetic environment.
Defining ${\rho}=(\rho_{11},\rho_{12},\rho_{21},\rho_{22})$, one obtains that it satisfies
the Liouville--von Neumann equation $\dot\rho=\mathcal{L} \rho$ associated with the superoperator
\begin{align}
\mathcal{L} = 
\left( \begin{array}{cccc} 
0 & i\Omega_{\rm L}/2 & -i\Omega_{\rm L}/2 & \Gamma \\
i\Omega_{\rm L}/2 & -i\delta-\Gamma/2 & 0 & -i\Omega_{\rm L}/2 \\
-i\Omega_{\rm L}/2 & 0 & i\delta-\Gamma/2 & i\Omega_{\rm L}/2 \\
0 & -i\Omega_{\rm L}/2 & i\Omega_{\rm L}/2 & -\Gamma \\
\end{array} \right)
\end{align}
in the rotating wave approximation, where $\delta=\omega_{\rm L}-\omega_{\rm TLS}$ defines the laser detuning.
This corresponds to the optical Bloch equations that are known to provide a realistic description 
of electric dipoles in single-molecule experiments \cite{lounis1997single,brunel1998rabi}.
In the stationary regime $\mathcal{L} \rho_{0}$=0, the quantum regression theorem leads to the autocorrelations
\begin{align}\label{CorrelationFuncion}
    S(t)
    &= \langle \sigma_z(t)\,\sigma_z(0) \rangle
    - \langle \sigma_z \rangle^2 \notag \\
    &=\langle w_{0} | \mathcal{M}_{z}e^{\mathcal{L}t}\mathcal{M}_{z} | \rho_{0} \rangle - \langle w_{0} | \mathcal{M}_{z} |\rho_{0}\rangle^{2} \,,
    \end{align}
where $w_{0}=(1,0,0,1)$ denotes the kernel left-hand eigenvector of $\mathcal{L}$, and $\mathcal{M}_{z}=\Diag(1,1,-1,-1)$ 
(see, also, Ref. \onlinecite{pistolesi2018bistability}).
The power spectral density $S(\omega)=\int dt\, e^{i\omega t}\, S(t)$ characterizes the absorption ($\omega>0$) and emission
 ($\omega<0$) of the TLS.
The frequency asymmetry of the quantum noise relates to the imaginary part of the response function through $2\Imag S_{\rm R}(\omega)=S(-\omega)-S(\omega)$.
We then obtain
\begin{align}\label{Autocorrelation}
S_{\rm R}(\omega) = \frac{8\delta\Omega_{\rm L}^{2}}{\Gamma^{2}+(2\delta)^{2}+2\Omega_{\rm L}^{2}} \frac{i\omega-\Gamma}{P(-i\omega)} \,,
\end{align}
where the polynomial $P(z)=\sum_{n=0}^{3}a_{n}\,z^{n}$ has real coefficients
$a_{0} = \Gamma(\Gamma^{2} + (2\delta)^{2} + 2\Omega_{\rm L}^{2})/4$, 
$a_{1} = (5\Gamma^{2} + (2\delta)^{2} + 4\Omega_{\rm L}^{2})/4$,
$a_{2} = 2\Gamma$, and
$a_{3} = 1$.
Figure\,\ref{PowerSpectrum} shows the coupling strength $S_{\rm R}\equiv S_{\rm R}(\omega_{0})$ as a function of the laser detuning and the 
Rabi frequency.
%
%
The coupling vanishes when the coherent drive is in resonance ($\delta=0$) or strongly detuned ($|\delta| \gg\Gamma$) with respect to the splitting of the TLS.
It also vanishes when the coherent drive is turned off ($\Omega_{\rm L}=0$) or when it is sufficiently strong to 
saturate the populations of the TLS ($\Omega_{\rm L}\gg\Gamma$). 
In these situations, the two eigenstates are oscillations along two orthogonal directions.
Otherwise, the TLS mediates an effective coupling that modifies the non-Hermitian dynamics of the mechanical modes 
leading to complex eigenvalues $\lambda_{\pm}$ and eigenstates $\bf X_{\pm}$ of $H$.
In the next section, we show how degeneracies in the complex spectrum of $H$ are associated with singular properties of the eigenstates 
that affect the polarization of the mechanical modes in real space.

\section{Exceptional degeneracy points}

\subsection{Electromechanical spectrum}
For nonvanishing coupling ($g_1 g_2 S_{\rm R}\neq 0$), the electromechanical spectrum may exhibit exceptional 
degeneracy points in the parameter space $(\delta,\Omega_{\rm L})$, which describes the laser driving.
They occur when the retarded TLS response satisfies
\begin{align}\label{Critical SR}
S_{\rm R}=S_{\rm EP\pm} \equiv - \frac{\Delta\omega}{G_{\pm}} \,,
\end{align}
where $\Delta\omega=\omega_{1} - \omega_{2} - i(\gamma_{1}-\gamma_{2})/2$ characterizes the splitting of the mechanical modes and $G_\pm = -(g_{1}^{2}-g_{2}^{2} \pm i 2 g_{1}g_{2})$ relates to the coupling strengths between the mechanical oscillators and the TLS. 
We focus in particular on the degeneracy point $S_{\rm EP}\equiv S_{\rm EP-}$, which lies in the positive imaginary plane.
It is then convenient to chose this EP as a new origin of the complex plane, such that $S_{\rm R}=S_{\rm EP}+z$.
The mechanical dynamics is described equivalently in terms of an effective Hamiltonian $H'$ similar to $H$ 
(see, also, Appendix\,\ref{Appendix Exceptional point properties}).
We find
\begin{equation}\label{JordanForm}
    H\sim H' = \frac{1}{2}
    \left(
    \begin{array}{cc}
       h_0 & 0 \\
       G_+\Delta S & h_0
    \end{array}
    \right)
    +
    \frac{z}{2}
    \left(
    \begin{array}{cc}
       0 & G_- \\
       G_+ & 0
    \end{array}
    \right)\,,
\end{equation}
where $\Delta S\equiv S_{\rm EP-} -S_{\rm EP +}$ depends on the distance between the two EPs and $h_{0}= \omega_{1}+\omega_{2} -i (\gamma_{1}+\gamma_{2})/2-(g_{1}^{2}+g_{2}^{2}) \,S_{\rm R} $.
This representation explicitly shows that the effective Hamiltonian $H$ supports a Jordan matrix representation at the EP in $z=0$.

\begin{figure*}
    \centering
    \includegraphics[trim = 0mm 128mm 290mm 0mm, clip, width=18cm]{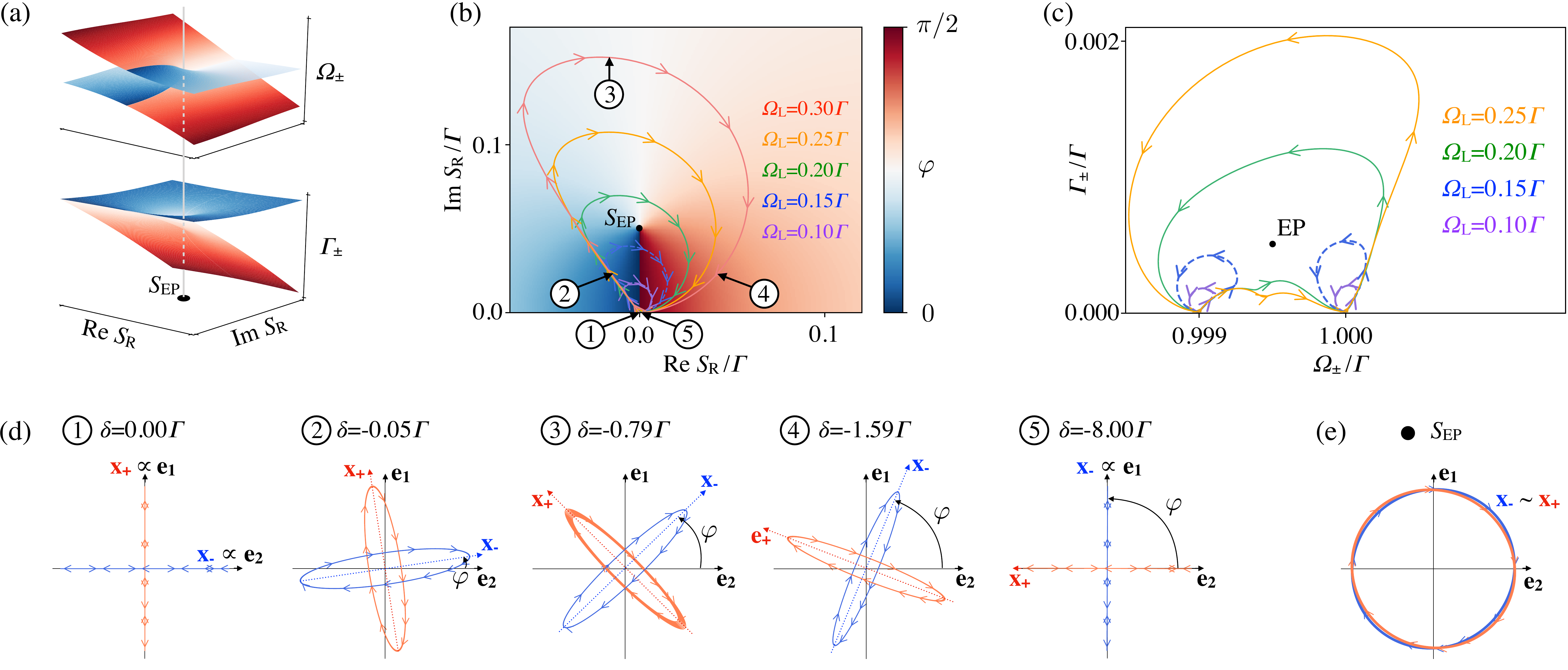}
    \caption{\textbf{Mean-field mechanical spectrum and eigenmode polarization.}
    The parameters are $\omega_1=\Gamma$ and $\omega_{2}=\omega_1-10^{-3}\Gamma$, 
    $\gamma_i=10^{-5}\Gamma$, and $g_i=0.1\Gamma$.
    (a) Riemann-surface representation of the mechanical spectrum near the exceptional point $S_{\rm EP}$, where the resonance frequencies $\Omega_{\pm}$ and damping rates $\Gamma_{\pm}$ of the mechanical eigenmodes are degenerate.
    (b) Oriented loops of the retarded TLS response $S_{\rm R}$ when varying the detuning $\delta$ from 0 to $-8\Gamma$ for various Rabi frequencies $\Omega_{\rm L}$.
    (c) Evolution of the eigenvalues $\lambda_{\pm}$ in the complex plane when $S_{\rm R}$ varies along the loops in (b). The loops enclosing the exceptional point $S_{\rm EP}$ lead to the switch of the mechanical frequencies $\omega_{1,2}$ and damping rates $\gamma_{1,2}$.
    (d) Oscillations of the eigenmodes ${\bf x_{\pm}}$ over a time scale of $50/\Gamma$ for various values of $\delta$ along the red loop $\Omega_{\rm L}=0.3\Gamma$ in (b). The elliptical polarization of the oscillations undergoes a roation of angle $\varphi$, which is specified by the colormap in (b).
    (e) Oscillations of the eigenmodes ${\bf x_{\pm}}$ near the EP over a time scale of $50/\Gamma$, for $\delta=-0.85\Gamma$ and $\Omega_{\rm L}=0.17\Gamma$.
    }
    \label{MeanFieldFig}
\end{figure*}

The electromechanical spectrum presents the two eigenvalues
\begin{equation}
    \lambda_\pm = \frac{h_0}{2}\pm\frac{1}{2}\sqrt{G_+G_-(\Delta S+z)z} \,.
\end{equation}
Therefore, the EP in $z=0$ also corresponds to the branch point of the complex square root $\sqrt{z}$.
The resonance frequencies $\Omega_{\pm}=\Real [\lambda_{\pm}]$ and the damping rates $\Gamma_{\pm}=-\Imag [\lambda_{\pm}]$ then support a Riemann-surface representation in the vicinity of the EP [Fig.\,\ref{MeanFieldFig}(a)].
This can be evidenced by varying the detuning $\delta$ and Rabi frequency $\Omega_{\rm L}$ of the TLS drive.
%

%
Figure\,\ref{MeanFieldFig}(b) shows that strongly detuning the TLS drive from the resonance ($\delta=0$) outlines a loop in parameter space.
The response function goes away from and back to $S_{\rm R}=0$, where the mechanical coupling mediated by the TLS vanishes (Fig.~\ref{PowerSpectrum}).
This point corresponds to the bare mechanical frequencies and damping rates, which are $\Omega_{\pm}=\omega_{1,2}$ and $\Gamma_{\pm}=\gamma_{1,2}$.
One can check in Fig.\,\ref{MeanFieldFig}(c) that the mechanical frequencies and damping rates go back to their bare initial values when the loop does not enclose the EP in parameter space.
For EP-enclosing loops, however, the mechanical frequencies $\Omega_{\pm}$ and damping rates $\Gamma_{\pm}$ do not come back on their initial values, 
but are exchanged.
This eigenvalue switch is an evidence of the Riemann-surface topology in the electromechanical spectrum.
For the case of an optical cavity coupled to two eigenmodes of a membrane, this effect has been observed in Ref.~\onlinecite{xu2016topological}.
Here we showed that one can have a similar behavior for a TLS.
Specifically, for the electromechanical system that we propose in Fig.~\ref{fig:system}, the condition of existence for the EP in the positive imaginary plane reads $S_{\rm EP}\simeq i\Delta\omega/(2g_{i}^2) $, assuming $g_1\simeq g_2$.
Since $S_{\rm EP}$ is nearly pure imaginary in the underdamped regime ($\gamma_{i} \ll \omega_{i}$), varying the TLS response $S_{\rm R}$ around this degeneracy point requires ${\rm Re}[S_{\rm R}]$ to change signs and  $|S_{\rm R}|\simeq|S_{\rm EP}|$.
These two conditions give $\omega_{i}>\Gamma/2$ and $\Gamma\Delta\omega\simeq2g_{i}^2$, for the maximum of $S_{\rm R}$ is obtained when $\delta\sim\Omega_{\rm L}\sim\Gamma$ (Fig.\,\ref{PowerSpectrum}).
For the TLS of electric dipoles recently observed \cite{Moradi2019Matrix}, experiencing the EP then implies a typical coupling $g_i/(2\pi)\approx 0.3$\,MHz.
It is much smaller than the critical coupling $g_{c}\simeq10^3$ MHz and could be realized, for instance, by positioning the tip of a carbon nanotube 100\,nm away from the molecule with a $100$\,$\mu$V bias.
We emphasize that such a coupling is not strong enough to excite higher flexural modes.
According to Euler-Bernoulli beam theory \cite{cleland2013foundations}, their frequencies are at least six times larger than the fundamental ones.
%

\begin{figure*}[thb]
    \centering
    \includegraphics[trim = 0mm 88mm 33mm 0mm, clip, width=18
    cm]{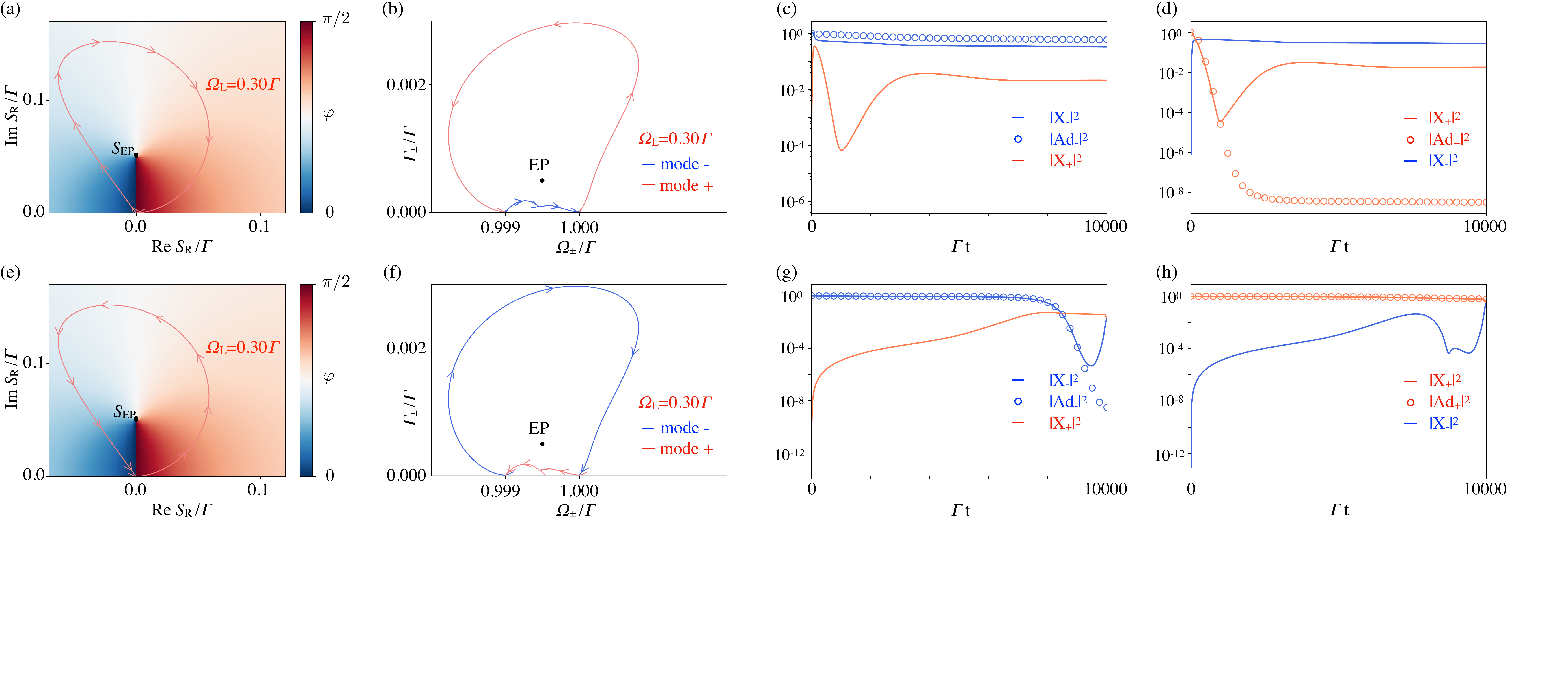}
    \caption{\textbf{Chiral nature of the polarization around the EP.}
    The parameters are
$\omega_1=\Gamma$ and $\omega_{2}=\omega_1-10^{-3}\Gamma$, $\gamma_i=10^{-5} \Gamma$, and $g_i=0.1\Gamma$.
    (a) Clockwise-oriented loop obtained by varying $\delta$ from 0 to $-8\Gamma$. 
    (b) Evolution of the eigenvalues along the loop in (a). 
    (c) Comparison between the exact and adiabatic time evolutions when ${\bf X}(0)={\bf X_-}(0)={\bf X_2}$. 
    (d) Comparison between the exact and adiabatic time evolutions when ${\bf X}(0)={\bf X_+}(0)={\bf X_1}$ over a time scale $T=10^{5}\Gamma$. 
    (e) -- (h) Same as the first row, but for a counterclockwise-oriented loop obtained by varying $\delta$ from $-8\Gamma$ to 0.}
    \label{Adiabaticity}
\end{figure*}

\subsection{Eigenmode polarization around the EP}
The detailed analysis of the eigenvectors unveils a very interesting dynamics of the oscillator tip.
%
%
To investigate them, we choose the biorthogonal left and right eigenstates of the effective Hamiltonian $H'$ as 
\begin{equation}\label{Eigenstates0}
{\bf Y_{\pm}} = \frac{1}{\sqrt{2}}
\left( \begin{array}{cc} 
\pm Z^{\frac{1}{4}} \,; &
Z^{-\frac{1}{4}} \\
\end{array} \right)
 ~~~\text{and}~~~ {\bf X_{\pm}} = \frac{1}{\sqrt{2}}
\left( \begin{array}{c} 
\pm Z^{-\frac{1}{4}} \\
Z^{\frac{1}{4}} \\
\end{array} \right) ,
\end{equation}
where we use the polar representation $z=\rho e^{i\theta}$ around the EP, so that
\begin{align}\label{PolarZ}
Z(\theta)=\frac{G_{+}}{G_{-}}\left(\frac{\Delta S}{\rho} e^{-i\theta} + 1\right) .
\end{align}
The exceptional degeneracy point is associated with a phase singularity for the eigenstates, which can also be evidenced by varying $S_{\rm R}$ smoothly around the degeneracy.
For a $\rho$-radius loop that encloses only $S_{\rm EP}$ ($\rho<|\Delta S|$), the eigenstates fulfill the condition of parallel transport ${\bf Y_{\pm}}(\theta)\cdot \nabla_\theta {\bf X_{\pm}}(\theta)=0$ and are multivalued.
They change as ${\bf X}_{\pm}(\theta$+$2n\pi) = (-i)^{n} {\bf X}_{\pm}(\theta)$ if $n$ is even, and ${\bf X}_{\pm}(\theta$+$2n\pi) = (-i)^{n} {\bf X}_{\mp}(\theta)$ if $n$ is odd.
Therefore, additionally to the eigenvalues, the eigenstates also switch after one loop around the EP.
The multivaluation of the eigenstates further affects the polarization of the mechanical oscillations ${\bf x_{\pm}}(t)$ in real space.
The eigenstates are solutions of the effective Schr\"odinger equation, that is, ${\bf X_{\pm}}(t)=x_{\pm}(0)\,e^{-i\lambda_{\pm}t}{\bf X_{\pm}}$, where $x_{\pm}(0)$ are initial amplitudes that we assume to be real.
The mechanical dynamics then consists of damped oscillations with {\em elliptical} polarization (see, for more details, Appendix\,\ref{Appendix Exceptional point properties}).
We find
\begin{align}\label{Rotation}
{\bf x}_{\pm}(t) = R(\varphi)\, x_{\pm}(0)\,e^{-\Gamma_{\pm}t} 
\left( \begin{array}{c}
\alpha\,\cos(\Omega_{\pm}t) \\
\beta\,\sin(\Omega_{\pm}t) \\
\end{array} \right) .
\end{align}
The semiaxes $\alpha$ and $\beta$ of the ellipse determine the eccentricity of the mechanical oscillations, and $R(\varphi)= \tau_0 \cos\varphi - i\tau_2 \sin\varphi$ is a rotation matrix where the Pauli matrices $\tau_{i}$ are written in the orthogonal basis (${\bf e_{1}}$,${\bf e_{2}}$) of the uncoupled modes.
The rotation angle $\varphi$ thus characterizes the orientation of the polarization with respect to one of the modes in the absence of coupling ($g_{1}g_{2}S_{R}=0$).
Figure\,\ref{MeanFieldFig}(d) presents the dynamics of the mechanical oscillations when the retarded response of the driven TLS performs a loop around the EP in parameter space.
The semiaxes of the elliptical polarization remain unchanged after the loop.
We find that this property generally holds for any loop since the semiaxes vary as $\alpha(\theta+2\pi)=\alpha(\theta)$ and $\beta(\theta+2\pi)=\beta(\theta)$ (Appendix\,\ref{Appendix Exceptional point properties}).
Nevertheless, the mechanical oscillations do not go back to the initial polarization at the end of the loop, for the polarization undergoes a rotation of $\varphi=\pi/2$.
We can more generally show that the polarization rotates as $\varphi(\theta-2n\pi)=\varphi(\theta)+n\pi/2$ (Appendix\,\ref{Appendix Exceptional point properties}).
This fourfold invariance is reminiscent of the fourth-root multivaluation of the eigenstates in Eq.\,(\ref{Eigenstates0}).
We emphasize that the $\pi/2$ rotation of the polarization also comes with the switch of the resonance frequencies and damping rates.
Thus, an eigenmode initially activated as 
${\bf x}(t) = x(0)e^{-\gamma_{2}t}\cos(\omega_{2}t)\,{\bf e_{2}}$ is transferred into the eigenmode of orthogonal polarization, that is,
${\bf x}(t) \propto x(0)e^{-\gamma_{1}t}\cos(\omega_{1}t+\Delta \phi)\,{\bf e_{1}}$, 
where $\Delta \phi$ is the phase accumulated along the EP-enclosing loop. 
This transfer of energy from one eigenmode to the other only depends on whether $S_{\rm R}$ encircles $S_{\rm EP}$, regardless of the precise loop geometry.
The actuation between mechanical modes is therefore topological.

\subsection{Chiral nature of the polarization}

The topological actuation from one mode to the other is an intrinsic property of the instantaneous eigenstates ${\bf X_{\pm}}$.
To observe the effects of their multivaluation around the EP, we further investigate their adiabatic transport.
In Hermitian systems, the adiabatic theorem ensures that one can neglect the nonadiabatic transitions over some typical timescale $T \gg 1/|\lambda_+-\lambda_-|$.
In open systems, however, this is no longer true for all the eigenstates \cite{uzdin2011observability,berry2011slow,xu2016topological}.
For a two-state system, in particular, only the least dissipative state is expected to be transported adiabatically around the EP \cite{milburn2015general}.
Here we study this issue by solving numerically the Schr\"{o}dinger-like equation $i\dot {\bf X}=H {\bf X}$.
We focus on the EP-enclosing loop associated with the Rabi frequency $\Omega_{\rm L}=0.3\Gamma$ in Fig.\,\ref{MeanFieldFig}(b).
We ramp the detuning linearly over the time scale $T=10^{5}/\Gamma$ between $\delta=0$ and $\delta=-8\Gamma$.
Then, $|{\bf X_{\pm}}(t)|^2=|{\bf Y_{\pm}}(t)\mathcal{U}(t,0){\bf X}(0)|^2$ provides the exact dynamics, where $\mathcal{U}$ denotes the time-evolution operator for the Hamiltonian $H$.
We assume the initial eigenstates are either ${\bf X}(0)={\bf X_+}={\bf X_1}$ or ${\bf X}(0)={\bf X_-}={\bf X_2}$ associated with the frequencies
$\omega_1=\Gamma$ and $\omega_{2}=\omega_1-10^{-3}\Gamma$.
We then compare the exact dynamics to their adiabatic evolutions $|Ad_{\pm}(t)|^2=|\exp{[-i\int_0^t \lambda_\pm(\tau) d\tau]} {\bf Y_{\pm}}(0)\,{\bf X}(0)|^2$.
The results are presented in Fig.\,\ref{Adiabaticity} for two orientations of the EP-enclosing loop in parameter space.
For a clockwise loop [Fig.\,\ref{Adiabaticity}(a)], the eigenstate ${\bf X_-}$ is the least dissipative one [Fig.\,\ref{Adiabaticity}(b)]
\begin{equation}
    \int_{0}^{T}dt \left[\Gamma_{+}(t)-\Gamma_{-}(t) \right]>0 \,.
\end{equation}
We find that only this state can experience the adiabatic transport [Figs.\,\ref{Adiabaticity}(c) and (d)].
When reversing the orientation of the loop, the situation is reversed too.
The eigenstate ${\bf X_+}$ becomes the least dissipative one and follows the adiabatic evolution, whereas ${\bf X_-}$ does not (Figs.\,\ref{Adiabaticity}e-h).
Thus, both modes ${\bf x_{\pm}}$ can experience the topological actuation, but for opposite orientations of the loop.
This asymmetry with respect to the orientation of the loop reveals the chiral nature of the $\pi/2$ rotation of the eigenmode polarization around the EP.
%

\begin{figure*}[t]
    \centering
    \includegraphics[trim = 0mm 232mm 245mm 0mm, clip, width=18cm]{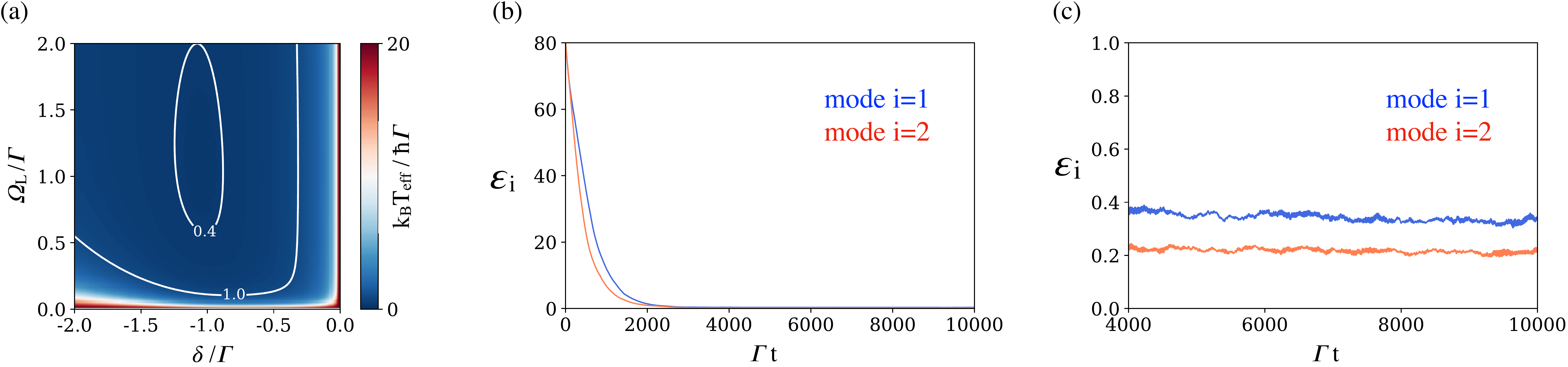}
    \caption{\textbf{Mechanical cooling via the TLS noise.}
    The parameters are $\omega_1=1.000\Gamma$ and $\omega_{2}=0.999\Gamma$, $\gamma_i=0.00001\Gamma$, and $g_i=0.1\Gamma$.
    (a) Effective temperature as a function of the detuning $\delta$ and Rabi frequency $\Omega_{\rm L}$.
    (b) Evolution of the dimensionless mechanical energies $\mathcal{E}_i$ during a time period of $10000/\Gamma$, obtained from quantum-jump Monte Carlo simulations averaged over 1000 dynamics for $\delta=-\Gamma$ and $\Omega_{\rm L}=\Gamma$. 
    (c) Zoom of (b) showing that the two modes can be cooled below $\hbar \Gamma$ ($\simeq0.5$\,mK) via their coupling to the TLS.}
    \label{Teff}
\end{figure*}

\subsection{Eigenmode polarization at the EP}
We can also investigate the fate of the mechanical oscillations when approaching the EP.
Figure\,\ref{MeanFieldFig}(e) illustrates this situation.
We find that the two mechanical modes coalesce into a single mode of circular polarization.
The coalescence results from the eigenstates that become collinear in the limit of small $\rho$ in Eq.\,(\ref{Eigenstates0}).
The circular polarization is reminiscent of the Jordan matrix representation of $H$ and is then a manifestation of the EP.
Nevertheless, it is a local property in parameter space for which noise may be detrimental, especially because noise is known to be enhanced by the extreme nonorthogonality of the eigenstates close to the EP \cite{mortensen2018fluctuations,lau2018fundamental,zhang2019quantum,naghiloo2019quantum}.
Another drawback is that the EP is a degeneracy point where, by definition, the band gap closes and around which nonadiabatic transitions then become unavoidable \cite{Hassan:2017aa}.
In contrast, the $\pi/2$ rotation of the polarization that we have introduced above evidences a global (topological) property of the EP for which the adiabatic transport is possible.
It does not require to approach the EP and, therefore, should be a more robust manifestation.

\section{Quantum noise and back-action}

\subsection{TLS quantum fluctuations}
The mean-field picture discussed above is appealing, for it introduces a topological property of the EP to manipulate the vectorial polarization of mechanical oscillations.
Now, we show that this manifestation of the EP can also survive the fluctuations of the open quantum system that dresses the mechanical modes. 
In optical cavities, the number of circulating photons $n_{c}$ obeys a Poisson distribution, and so $\langle \delta n_{c}^{2} \rangle=\langle n_{c} \rangle$.
The fluctuations of the radiation pressure force $F$ then satisfy $\langle \delta F^{2} \rangle/\langle F \rangle^{2}=1/\langle n_{c} \rangle$, which becomes negligible for the usual cavities that are strongly populated.
For the TLS, however, the force scales as $F\propto \hat n_{g}-\hat n_{e}$, where the populations of the ground and excited states satisfy the constraints $\hat n_{g}+ \hat n_{e}=1$ and $\hat n_{g} - \hat n_{e}=\sigma_{z}$.
%
%
%
%
The force fluctuations verify
\begin{align}
\frac{\langle \delta F^{2} \rangle}{\langle F \rangle^{2}}=\frac{\langle \hat n_{e} \rangle(1-\langle \hat n_{e} \rangle)}{1/4-\langle \hat n_{e} \rangle(1-\langle \hat n_{e} \rangle)} \,,
\end{align}
where we used $\langle \hat n_{e}^{2} \rangle=\langle \hat n_{e} \rangle$.
Since $\langle \hat n_{e} \rangle\leq1/2$, the fluctuations can be arbitrarily large compared to the mean force, and eventually $\langle \delta F^{2} \rangle/\langle F \rangle^{2}\rightarrow\infty$.
To describe the TLS force fluctuations due to spontaneous emission and test the mean-field description based on Eq.\,(\ref{ClassicalEQM}), we perform 
quantum-jump Monte Carlo simulations
\cite{gardiner1991quantum,dalibard1992wave,hekking2013quantum}.
This implies solving, on short timescales $\Delta t\ll 1/\Gamma$, the differential equations
\begin{align}\label{QuantumJumpEquations}
i \dot a_{\rm g} &=   
\delta[{\bf x}(t)] a_{\rm g}/2 + \Omega_{\rm L}a_{\rm e}/2 \\
i \dot a_{\rm e} &= \left(\delta[{\bf x}(t)]/2-i\Gamma\right) a_{\rm e} + \Omega_{\rm L}a_{\rm e}/2 \notag \\
m\ddot{x}_i &= -m\omega_i^2 x_i-m\gamma_i\dot{x}_i + 
\delta F_{\rm th} +\frac{g_i}{x_i^{\rm zpf}}\langle\sigma_z\rangle \notag \,,
\end{align} 
where we introduce $\delta[{\bf x}(t)]=\delta-\sum_i g_i x_i(t)/x_i^{\rm zpf}$, the Brownian thermal force $\delta F_{\rm th}$ with variance $\sqrt{2m\gamma_i k_B T_i}$,
and the wavefunction of the TLS, $|\psi(t)\rangle=a_{\rm g}(t)|{\rm g}\rangle+a_{\rm e}(t)|{\rm e}\rangle$, so that
$\langle \sigma_z(t)\rangle=|a_{\rm g}(t)|^2-|a_{\rm e}(t)|^2$.
At each time step, one randomly either allows a transition to the ground state with probability 
$\Gamma \Delta t |a_e|^2$, or normalizes the wavefunction and proceeds with the time evolution. 
%

\begin{figure*}
    \centering
    \includegraphics[trim = 0mm 252mm 258mm 1mm, clip, width=18cm]{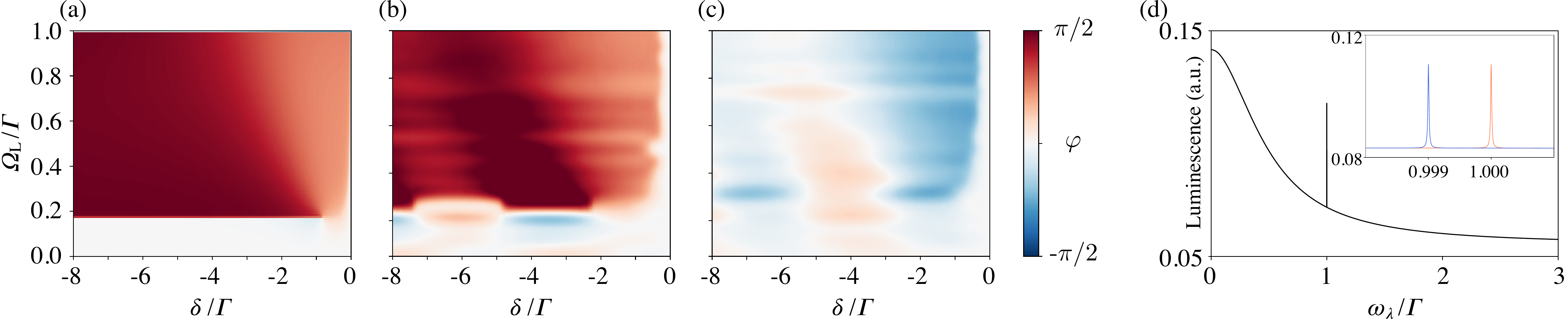}
    \caption{\textbf{Noise-averaged rotation of the mechanical polarization.}
    The parameters are $\omega_1=1.000\Gamma$ and $\omega_{2}=0.999\Gamma$, $\gamma_i=0.00001\Gamma$, and $g_i=0.1\Gamma$.
    (a) Rotation angle $\varphi$ of the mechanical polarization as a function of the laser detuning $\delta$ and Rabi frequency $\Omega_{\rm L}$ obtained from the mean-field description.
    (b) Same as (a) but obtained from the quantum-jump Monte Carlo simulations averaged over 100 dynamics when the mode 2 is initially activated.
    (c) Same as (b) when the mode $1$ is initially activated. 
    (d) Luminescence excitation spectrum of the TLS driven by a coherent field of frequency modulation $\beta_\lambda=1$. The interference peaks, resolved in the inset, are centered on the mechanical frequencies $\omega_1=1.000\Gamma$ or $\omega_2=0.999\Gamma$, depending on the mode topologically activated at the end of the loop.}
    \label{MonteCarlo}
\end{figure*}

\subsection{Mechanical cooling via the TLS noise}
The mean-field description introduced above involves the response function of the bare TLS, that is, in the absence of electromechanical coupling ($g_{i}=0$).
This neglects the effects of the detuning shift $\delta[{\bf x}(t)]$ induced by the mechanical displacement.
In particular, large oscillations may lead to important effects when $g_i x_i/x_i^{\rm zpf} \sim \Gamma$, since the oscillations can effectively change the laser-TLS detuning.
%
It is then interesting to work at low temperature. 
We investigate here how the system can be cooled down by coupling to the TLS \cite{puller2013single}.

The power spectrum $S(\omega)$ of the TLS noise can be obtained from Eq.\,(\ref{CorrelationFuncion}). We denote its symmetric- and asymmetric-in-frequency parts, $S_\pm(\omega)=S(\omega)\pm S(-\omega)$.
The optical damping induced by the TLS noise relates to the asymmetry between emission and absorption as $\gamma_{\rm TLS}=g_0^2 S_-(\omega_0)$, where we assume $g_{1,2}= g_{0}$.
Cooling the mechanical oscillator via the TLS then requires the optical damping to be stronger than the intrinsic damping of the thermal baths, $\gamma_0\ll\gamma_{\rm TLS}$, where $\gamma_{1,2} = \gamma_{0}$.
The natural frequency scale of the noise must relate to the spontaneous emission rate $\Gamma$, so one can expect $S_{-}=-2\Imag S_{\rm R}\sim 1/\Gamma$.
One can check that this is indeed the case for $\delta=\Omega_{L}=\Gamma=\omega_{0}$ in Eq.\,(\ref{Autocorrelation}).
As the coupling cannot be larger than the critical one of the discharge electric field between the TLS and the carbon nanotube, 
we find that mechanical cooling is possible when
\begin{align}
    \omega_0\gamma_0 \ll g_0^2 < g_c^2 \,.
\end{align}
For the electromechanical system in Fig.\,\ref{fig:system}, one has $\omega_0\gamma_0$ that is at least three orders of magnitude smaller than the critical coupling $g_c$, which leads to a large range of possible coupling strength $g_{0}$.
Besides, the TLS temperature verifies $2k_{\rm B} T_{\rm TLS}/(\hbar \omega_0) = S_+/S_-$.
This leads to the effective temperature $T_{\rm eff}$ for the mechanical modes dressed by the TLS,
\begin{align}\label{EffTemp}
    \frac{2k_{\rm B} T_{\rm eff}}{\hbar \omega_0}
    = \frac{\gamma_0\frac{2k_{\rm B} T_{0}}{\hbar \omega_0}+\gamma_{\rm TLS}\frac{2k_{\rm B} T_{\rm TLS}}{\hbar \omega_0}}{\gamma_0+\gamma_{\rm TLS}} \,.
\end{align}
Figure\,\ref{Teff}\,(a) represents the map of the effective temperature as a function of TLS drive parameters $\delta$ and $\Omega_{\rm L}$.
It shows that the mechanical system can be  cooled below $\hbar\Gamma\,(\simeq0.5$\,mK).
We further verify this possibility of mechanical cooling by means of quantum-jump Monte Carlo simulations.
Figure\,\ref{Teff}(b) presents the evolution of the mechanical energies averaged over 1000 dynamics for $\delta=-\Gamma$ and $\Omega_{\rm L}=\Gamma$, which corresponds to the region of maximum cooling predicted in Eq.\,(\ref{EffTemp}) and Fig.\,\ref{Teff}(a).
We study the mechanical energy through the dimensionless parameter $\mathcal{E}_{i}$ defined as
\begin{align}
\mathcal{E}_i=\frac{4\omega_{i}}{\Gamma} \frac{1}{\hbar \Gamma}\left(\frac{p_i^2}{2m}+\frac{1}{2}m \omega_i^2 x_i^2\right) \,.
\end{align}
Each mechanical mode is initially assumed to be in thermal equilibrium with a bath of typical dilution-fridge temperature $T_{0}=10$ mK.
Thus, we begin with randomly generating the position and momentum of each mode according to a Boltzmann distribution.
The equipartition theorem implies that the initial mean mechanical energy in the figures corresponds to $\mathcal{E}_i=\frac{4\omega_n}{\Gamma}\frac{k_{\rm B}T_n}{\hbar\Gamma}=80$.
We then simulate the stochastic dynamics based on Eq.\,(\ref{QuantumJumpEquations}) with the usual parameters 
$\omega_1=1\Gamma$, $\omega_{2}=\omega_1-10^{-3}\Gamma$, $\gamma_i=10^{-5}\Gamma$, and $g_{i}=0.1\Gamma$.
Figure\,\ref{Teff}(c) shows from the mean dynamics that the system can be cooled down $\hbar\Gamma$ ($\sim0.5$\,mK), to about $50$\,$\mu$K, in agreement the mechanical cooling predicted in Fig.\,\ref{Teff}(a). 

\subsection{Mean rotation of the mechanical polarization}
We come to the main question we want to address with the simulations: does the EP remain stable in the presence of 
fluctuations and backaction?
To test the mean-field manifestation of the EP, we study the evolution of the mechanical mode 1 or 2 with initial amplitude $10x^{\rm zpf}$ and ramp adiabatically $\delta$ from 0 to -8$\Gamma$ over the timescale $T=10^{4}/\Gamma$ for various values of $\Omega_{\rm L}$.
Such an initial actuation can be achieved by forcing the charged nanotube tip with a transverse electric field oscillating at a frequency tuned on the selected mode frequency.
The mean dynamics of the oscillator, averaged over $10^2$ simulated trajectories, consists of quasiperiodic elliptical oscillations in real space.
We thus identify the semiaxes in each quasiperiod and determine their rotation angle $\varphi$ at a given value of $\delta$ and $\Omega_{\rm L}$.
Figure\,\ref{MonteCarlo}(a) presents the prediction of the mean-field description, clearly showing the presence of the EP at $\delta\simeq-0.85\Gamma$.
In comparison, Fig.\,\ref{MonteCarlo}(b) shows the rotation angle for the same parameters, but obtained from the Monte Carlo
simulations. 
Though fluctuations seem to blur the EP, where the noise is enhanced by the coalescence of the nonorthogonal eigenstates \cite{mortensen2018fluctuations,lau2018fundamental,zhang2019quantum,naghiloo2019quantum}, the axis of the oscillations performs on average a rotation in agreement with the mean-field description.
Starting with an oscillation in the horizontal direction ($\varphi=0$), one ends with a perpendicular oscillation ($\varphi=\pi/2$), showing that the adiabatic picture remains valid.
If the rotation implies a transfer of energy from mode 1 to mode 2, continuously following the evolution of the oscillation axis proves that this transfer is purely due to the adiabatic evolution predicted by the mean-field approach, and not to dissipation or stochastic effects.
Figure\,\ref{MonteCarlo}(c) shows the same evolution from mode 1.
No energy transfer is realized, as expected from the mean-field adiabatic transport.
We would like to stress that the numerical calculation fully takes into account the nonlinearity and non-Gaussian behavior of the TLS, 
thus confirming the, at least approximate, validity of the mean-field approach.
In the next section, we finally discuss a possible experimental implementation for the direct detection of the energy transfer. 

\section{Detection via frequency modulations}

%
After the adiabatic loop in parameter space, the mechanical energy of one mode is transferred to 
the other one.
This implies that the mechanical oscillation frequency has changed, but the variation is extremely small.  
Here we show that detecting such a small change can be performed by modulation of the frequency of the coherent field that drives the TLS.

At the semiclassical level, the Hermitian Hamiltonian of the driven TLS  can be written as
\begin{align}
H_{\rm TLS}(t) = 
\left( \begin{array}{cc}
-\frac{\omega_{\rm TLS}}{2} -g_{i}\,x_{i}(t) & \Omega_{\rm L}\left( e^{iF(t)} + e^{-iF(t)} \right)/2 \\
\Omega_{\rm L}\left( e^{iF(t)} + e^{-iF(t)} \right)/2 & \frac{\omega_{\rm TLS}}{2} + g_{i}\,x_{i}(t) \\
\end{array} \right) \,,
\end{align}
where $F(t)= \int_{0}^{t}d\tau \, [\omega_{\rm L} + f(\tau)]$ and $f(t) = a_{\lambda}\cos(\omega_{\lambda}t+\phi_{\lambda})$ describes the frequency modulation around frequency $\omega_{\rm L}$. 
Let $\psi(t)$ be a wave function satisfying the time-dependent Shr\"odinger equation $i \dot \psi(t) = H_{\rm TLS}(t) \, \psi(t)$.
We can then introduce the gauge transformation $\phi(t) = U^{\dagger}(t) \psi(t)$ based on the unitary operator
\begin{align}
U = 
\left( \begin{array}{cc}
e^{iF(t)/2} & 0 \\
0 & e^{-iF(t)/2} \\
\end{array} \right) .
\end{align}
This leads to $i \dot \phi(t) = \left[ U^{\dagger}(t)H_{TLS}(t)U(t) - i U^{\dagger}(t)\dot U(t) \right] \phi(t)$, where the effective Hamiltonian is
\begin{align}
\tilde H_{\rm TLS}(t) = 
\left( \begin{array}{cc}
\delta(t)/2 -g_{i}\,x_{i}(t) & \Omega_{\rm L}/2 \\
\Omega_{\rm L}/2 & -\delta(t)/2 + g_{i}\,x_{i}(t) \\
\end{array} \right) ,
\end{align}
and $\delta(t)=\delta-(\beta_\lambda \omega_\lambda)\cos(\omega_\lambda t)-g_i x_i(t)$.
Thus, the frequency modulation induces an additional detuning shift, which adds up to the one of the mechanical oscillator in the modulated rotating frame.
The frequency modulation and the mechanical oscillations may then lead to interference when $\omega_\lambda=\omega_i$, since only the mode $i$ is exited at the end of the loop.
We can show, in particular, that the interference affects the excited-state population of the driven TLS, and so its
luminescence excitation spectrum (Appendix\,\ref{Appendix Frequency modulation of the TLS drive}).
This is illustrated in Fig.\,\ref{MonteCarlo}(d).
The Lorentzian background of width $\Gamma$ and centered on $\omega_{\lambda}=0$ already exists in the absence of mechanical mode ($g_{i=0}$) and, thus, is not due to any interference.
The interference, however, appears through a narrow Lorentzian peak of width $\gamma_{i}$ centered on the mechanical frequency $\omega_i$ in the figure.
The width of the peak verifies $\gamma_i\ll\omega_{1}-\omega_{2}$, so that the two quasidegenerate modes could be resolved clearly in the experiments.

\section*{Conclusion}
Manipulating mechanical systems at the nanometer scale is an important challenge of present research.
The possibility of using EP in the excitation spectrum to transfer energy from one mechanical mode to the other had been 
proposed and observed in the past for mechanical modes coupled to optical cavities. 
In this paper, we have shown that EP can be equally generated by coupling mechanical modes to TLSs. 
Specifically, we considered a concrete example of single molecules coupled to flexural modes of carbon nanotubes, for which we performed detailed simulations.
We have shown quite generally that the topological and chiral energy transfer is possible.
Remarkably, the prediction of the analytical mean-field theory is confirmed by the quantum-jump Monte Carlo approach.
This guarantees that even if a TLS is quite different from an optical cavity, since it is a strongly nonlinear quantum 
system and has strong quantum fluctuations, it can be used to manipulate a mechanical oscillator exploiting the EP. 

From a conceptual point of view, the flexural mode eigenstates allow one to understand, 
in a transparent way, the mechanism of formation of the EP. 
We find that the evolution from the standard orthogonal eigenvectors to the coincident eigenvectors at the EP is 
performed by evolving the eigenvectors into elliptical oscillations, which eventually become circular at the EP.
The energy transfer is then simply obtained by a rotation of the axis of the elliptic oscillation of $\pi/2$ when a loop is performed around the EP.
These findings can allow a manipulation of the tip of the nanotube without the addition of any external electric fields. 
This possibility of quantum manipulation could, for instance, find applications in vectorial force microscopy, where monitoring the eigenmode polarization is crucial to scan a surface.

\begin{acknowledgements}
We gratefully acknowledge {\em Conseil R\'egional de Nouvelle-Aquitaine} for financial support (Grant No. 2016-1R60306-00007470) and Idex Bordeaux (Maesim Risky project 2019  of the LAPHIA Program). B.L. acknowledges the Institut Universitaire de France.
\end{acknowledgements}

\newpage
\appendix
\onecolumngrid
\section{Derivation of the non-Hermitian Hamiltonian}
\label{Derivation of the non-Hermitian Hamiltonian}
The expectation value of the displacement operator obeys the Langevin equation (\ref{ClassicalEQM}). 
We begin with neglecting the fluctuating forces $\delta F_{i}$ and $\langle\hat\sigma_z\rangle_0$, which only shifts the equilibrium position of the oscillator.
In Fourier space, this leads to
\begin{align}\label{EoM Frequency}
m(-\omega^{2} - i\gamma_{i}\omega +\omega_{i}^{2}) x_{i}(\omega) &= \sum_{j=1,2} \frac{g_i}{x_i^{\rm zpf}}\frac{g_j}{x_j^{\rm zpf}} S_{\rm R}(\omega) x_{j}(\omega) \,.
\end{align}
Since we consider the underdamped regime ($\gamma_{i}\ll\omega_{i}$), the bare mechanical poles verify
$\omega_{i\pm} \simeq \pm \omega_{i} - i \gamma_i/2$.
We can linearize Eq.\,(\ref{EoM Frequency}) in the vicinity of the two mechanical frequencies $\omega_{i\pm}$.
The displacement can then be written as $x_i \simeq x_{i+}+x_{i-}$, where $x_{i+}$ ($x_{i-}$) describes the mechanical oscillations of positive (negative) frequency $\omega_{i+}$ ($\omega_{i-}$).
The positive- and negative-frequency oscillations satisfy:
\begin{align}
\left\{ \begin{array}{ll}
+i2m\omega_{i}(-i\omega) x_{i+}(\omega) = 2m\omega_{i}(\omega_{i} -i \frac{\gamma_{i}}{2}) x_{i+}(\omega) - \sum_{j=1,2} \frac{g_i}{x_i^{\rm z}}\frac{g_j}{x_j^{\rm z}} S_{\rm R} x_{j+}(\omega) \\
-i2m\omega_{i}(-i\omega) x_{i-}(\omega) = 2m\omega_{i}(\omega_{i} +i \frac{\gamma_{i}}{2}) x_{i-}(\omega) - \sum_{j=1,2} \frac{g_i}{x_i^{\rm z}}\frac{g_j}{x_j^{\rm z}} S_{\rm R}^{*} x_{j-}(\omega) \\
\end{array} \right. ,
\end{align}
where $S_{\rm R} \equiv S_{\rm R}(\omega_{0})$ and $\omega_{0}=(\omega_{1}+\omega_{2})/2$ is the mean mechanical frequency.
Since $x_i\simeq x_{i+}+x_{i-}=2{\rm Re}[x_{i+}]$, we only focus on the positive-frequency solution $x_{i+}$.
Fourier transforming back to the time domain finally results in
\begin{align}
-i\dot x_{i+}(t) &= \left(\omega_{i} -i \frac{\gamma_{i}}{2}\right) x_{i+}(t) + \sum_{j=1,2} g_i g_j S_{\rm R} x_{j+}(t) \,,
\end{align}
where we have used $2m\omega_i\simeq x_i^{\rm zpf}x_j^{\rm zpf}$.
Therefore, the positive-frequency oscillations of the mechanical modes are described by the vector ${\bf X}=(x_{1,+},x_{2,+})$.
It obeys the Shr\"odinger-like equation $i {\bf {\dot{X}}} = H\,{\bf X}$ with the non-Hermitian effective Hamiltonian,
\begin{align}
H &=
\left( \begin{array}{cc}
\omega_{1}-i \frac{\gamma_{1}}{2} - g_{1}^{2} S_{\rm R} & - g_{1} g_{2} S_{\rm R} \\
- g_{1} g_{2} S_{\rm R} & \omega_{2}-i \frac{\gamma_{2}}{2} - g_{2}^{2} S_{\rm R} \\
\end{array} \right) .
\end{align}
This relates to the mechanical displacement vector ${\bf x}=( x_{1}, x_{2})$ in real space as ${\bf x}$=$2\Real [ {\bf X} ]$,
\\

\section{Exceptional point properties}
\label{Appendix Exceptional point properties}

\subsection{Branch point in the spectrum}
We are now interested in the consequences of degeneracies in the non-Hermitian spectrum of $H$ on the dynamics of the mechanical modes.
To make them more explicit, we perform two subsequent $\pi/2$ rotations of axes $y$ and $z$ with respect to the Bloch sphere of $H$ eigenstates.
Thus, we introduce the unitary operator
\begin{align}
U =&
e^{-i\frac{\pi}{4}\tau_{y}} e^{-i\frac{\pi}{4}\tau_{z}}
=
\frac{1}{2}
\left( \begin{array}{cc}
1-i & -1-i \\
1-i & 1+i \\
\end{array} \right)
\end{align}
and perform the transformation $H' = U H U^{\dagger}$.
This results in $H' =\sum_{i} h_{i}\tau_{i}/2$ with
\begin{align}\label{Hamiltonian}
\left\{
\begin{array}{llll}
h_{0}&= \omega_{1}+\omega_{2} -i (\gamma_{1}+\gamma_{2})/2-(g_{1}^{2}+g_{2}^{2}) \,S_{\rm R} \\
h_{1}&=\omega_{1}-\omega_{2} -i (\gamma_{1}-\gamma_{2})/2 -(g_{1}^{2}-g_{2}^{2}) \,S_{\rm R} \\
h_{2}&=-2 g_{1}g_{2} \, S_{\rm R} \\
h_{3}&=0 
\end{array} \right.
\,.
\end{align}
Degeneracies in the electromechanical spectrum then occur when $S_{\rm R}=S_{\rm EP\pm}$, where
\begin{align}\label{Critical SR}
S_{\rm EP\pm} \equiv - \frac{\Delta\omega}{G_{\pm}} \,.
\end{align}
Here, $\Delta\omega=\omega_{1} - \omega_{2} - i(\gamma_{1}-\gamma_{2})/2$ characterizes the splitting of the mechanical modes and $G_\pm = -(g_{1}^{2}-g_{2}^{2} \pm i 2 g_{1}g_{2})$ relates to the coupling strengths between the mechanical oscillator and the TLS. 
We focus, in particular, on the degeneracy point $S_{\rm EP}\equiv S_{\rm EP-}$, which lies in the positive imaginary plane.
It is then convenient to set it as the new origin of the complex plane by introducing the variable $z=S_{\rm R}-S_{\rm EP}$.
Thus, we find that the effective Hamiltonian reads
\begin{equation}\label{JordanForm}
    H' = \frac{1}{2}
    \left(
    \begin{array}{cc}
       h_0 & 0 \\
       G_+\Delta S & h_0
    \end{array}
    \right)
    +
    \frac{z}{2}
    \left(
    \begin{array}{cc}
       0 & G_- \\
       G_+ & 0
    \end{array}
    \right)\,,
\end{equation}
where $\Delta S\equiv S_{\rm EP-} -S_{\rm EP +}$ depends on the distance between the two EPs.
This explicitly shows that the effective Hamiltonian $H$ supports a Jordan matrix representation at the EP in $z=0$.
The electromechanical spectrum relies on the two eigenvalues
\begin{equation}
    \lambda_\pm = \frac{h_0}{2}\pm\frac{1}{2}\sqrt{G_+G_-(\Delta S+z)z} \,.
\end{equation}
Therefore, the EP in $z=0$ also corresponds to the branch point of the complex square root $\sqrt{z}$, and hence the Riemann surface of the electromechanical spectrum in Fig.\,\ref{MeanFieldFig}(a).
The same features hold when focusing on the degeneracy point $S_{\rm EP+}$ in the negative imaginary plane.

\subsection{Eigenstate multivaluation}
The squareroot behavior of the spectrum near the EP also leads to singular properties for the eigenstates.
To investigate them, we choose the biorthogonal left and right instantaneous eigenstates of the effective Hamiltonian $H'$ as 
\begin{equation}\label{Eigenstates}
{\bf Y_{\pm}}(\theta) = \frac{1}{\sqrt{2}}
\left( \begin{array}{cc} 
\pm Z^{\frac{1}{4}} \,; &
Z^{-\frac{1}{4}} \\
\end{array} \right)
 ~~~\text{and}~~~ {\bf X_{\pm}}(\theta) = \frac{1}{\sqrt{2}}
\left( \begin{array}{c} 
\pm Z^{-\frac{1}{4}} \\
Z^{\frac{1}{4}} \\
\end{array} \right) \,.
\end{equation}
Using the polar representation $z=\rho e^{i\theta}$ around the EP, we find
\begin{align}\label{PolarZ}
Z(\theta)=\frac{G_{+}}{G_{-}}\left(\frac{\Delta S}{\rho} e^{-i\theta} + 1\right) \,.
\end{align}
Let us assume that $S_{\rm R}$ can be varied smoothly along a $\rho$-radius loop that encloses only $S_{\rm EP}$ ($\rho<|\Delta S|$).
Then, we find
\begin{align}\label{MultivaluationZ}
Z^{1/4}(\theta=2\pi)=-iZ^{1/4}(\theta=0) \,,
\end{align}
so that the eigenstates, which meet the condition of parallel transport ${\bf Y_{\pm}}\cdot \nabla_\theta {\bf X_{\pm}}=0$, change as follows:
\begin{equation}\label{Multivaluation}
    \left \{
    \begin{array}{llll}
    {\bf X_{\pm}}(2\pi) &= -i \, {\bf X_{\mp}}(0)  \\
    {\bf X_{\pm}}(4\pi) &= -1 \, {\bf X_{\pm}}(0)  \\
    {\bf X_{\pm}}(6\pi) &= +i \, {\bf X_{\mp}}(0)  \\
    {\bf X_{\pm}}(8\pi) &= +1 \, {\bf X_{\pm}}(0)
    \end{array}
    \right . \,.
\end{equation}
Thus, we expect the two eigenstates to swap after one loop, to pick up a geometrical Berry phase $\pi$ after two loops, and then come the eigenstates come back onto the initial states without any geometrical phase only after four loops.
The eigenstate swap comes from the multivaluation of the eigenstates around the EP.
The fourfold multivaluation results from the complex fourth root in the parallel-transported eigenstates in Eq.\,(\ref{Eigenstates}).
Note that if the loop encircles the two exceptional points ($\rho>|\Delta S|$), one can choose single-valued eigenstates along the loop, so that ${\bf X_{\pm}}(2\pi n) = {\bf X_{\pm}}(0)$ for any integer $n$.
The eigenstates neither pick a geometrical phase nor swap in this case.

\subsection{Eigenmode polarization}
The instantaneous eigenstates obey the Shr\"odinger-like equation $i\dot{\bf X}(t){\bf _{\pm}}=H'{\bf X_{\pm}}(t)$, and so ${\bf X_{\pm}}(t) = x_{\pm}(0) \, e^{-i\lambda_{\pm}t}{\bf X_{\pm}}$. 
In real space, the mechanical eigenmodes evolve in time as
\begin{align}
    x_{\pm}(t) &= 2 {\rm Re} \left[U^\dagger {\bf X_{\pm}}(t) \right] \notag \\
    &= x_\pm(0) {\rm Re}
    \left( \begin{array}{c}
    e^{-i\lambda_\pm t}e^{i\frac{\pi}{4}}(Z^{\frac{1}{4}}\pm Z^{-\frac{1}{4}}) \\
    -ie^{-i\lambda_\pm t}e^{i\frac{\pi}{4}}(Z^{\frac{1}{4}}\mp Z^{-\frac{1}{4}}) \\
    \end{array} \right) .
\end{align}
We can ignore the overall phase shift of $\pi/4$ and rewrite it as
\begin{align}\label{Eigenmodes}
    x_{\pm}(t) 
    &= x_\pm(0) \, e^{-\Gamma_\pm t}
    \left( \begin{array}{c}
    a_\pm \cos{(\Omega_\pm t+\phi_\pm)} \\
    a_\mp \sin{(\Omega_\pm t+\phi_\mp)} \\
    \end{array} \right) .
\end{align}
The amplitudes and the phases of the oscillations depend on the polar angle $\theta$ around the EP through
\begin{align}\label{SemiAxes}
    a_\pm &= |Z^{\frac{1}{4}}\pm Z^{-\frac{1}{4}}| \notag \\
    \phi_\pm &= {\rm Arg}[Z^{\frac{1}{4}}\pm Z^{-\frac{1}{4}}] \,,
\end{align}
where $Z$ has been introduced in Eq.\,(\ref{PolarZ}).
In particular, if $S_{\rm R}$ varies on a clockwise-oriented loop around $S_{\rm EP}$, the multivaluation in Eq.\,(\ref{MultivaluationZ}) requires the amplitudes and the phases of the oscillations to vary as
\begin{align}\label{AmpliAndPhase}
    a_\pm(-2\pi)&=a_\mp(0) \notag \\
    \phi_\pm(-2\pi) &= \phi_\mp(0) +\frac{\pi}{2} \,.
\end{align}

In addition, Eq.\,(\ref{Eigenmodes}) is the parametric equation of a rotated (damped) ellipse.
To make it explicit, we introduce a counterclockwise rotation matrix of angle $\varphi$, namely
\begin{equation}
    R(\varphi) =
    \left(
    \begin{array}{cc}
        \cos \varphi & -\sin\varphi \\
        \sin \varphi & \cos \phi
    \end{array}
    \right) ,
\end{equation}
so that
\begin{equation}\label{Rotated Ellipse}
    \left( \begin{array}{c}
    a_\pm \cos{(\Omega_\pm t+\phi_\pm)} \\
    a_\mp \sin{(\Omega_\pm t+\phi_\mp)} \\
    \end{array} \right)
    =
    R(\varphi)
    \left(
    \begin{array}{c}
        \alpha \cos(\omega_\pm t) \\
        \beta \sin(\omega_\pm t)
    \end{array}
    \right) .
\end{equation}
This straightforwardly leads to the following equalities:
\begin{align}\label{RotationEqualities}
    a_+ \cos(\phi_+) &= \alpha \cos(\varphi) \notag \\
    a_+ \sin(\phi_+) &= \beta \sin(\varphi) \notag \\
    a_- \cos(\phi_-) &= \beta \cos(\varphi) \notag \\
    a_- \sin(\phi_-) &= \alpha \sin(\varphi) \,.
\end{align}
Along with Eq.\,(\ref{AmpliAndPhase}), this implies that if $S_{\rm R}$ varies on a clockwise-oriented loop around $S_{\rm EP}$ from $\theta=0$ to $\theta=-2\pi$,
\begin{align}
    \alpha(-2\pi)\cos(\varphi(-2\pi))
    &= \alpha(0)\cos(\varphi(0)+\pi/2) \notag \\
    \alpha(-2\pi)\sin(\varphi(-2\pi))
    &= \alpha(0)\sin(\varphi(0)+\pi/2) \,,
\end{align}
and we find similar relations for $\beta$.
This demonstrates that the semiaxes $\alpha$ and $\beta$ have undergone a $\pi/2$ rotation at the end of the loop, albeit their lengths remain unchanged.

Therefore, the mechanical oscillations of the two eigenmodes have elliptical polarizations in real space,
\begin{align}
    x_{\pm}(t)
    &= x_\pm(0) \, e^{-\Gamma_\pm t}
    R(\varphi)
    \left(
    \begin{array}{c}
        \alpha \cos(\Omega_\pm t) \\
        \beta \sin(\Omega_\pm t)
    \end{array}
    \right) .
\end{align}
The $\varphi$ rotation of the polarization semiaxes $\alpha$ and $\beta$ is controlled by the polar angle $\theta$ around the EP in parameter space.
This behavior is illustrated in Fig.\,\ref{MeanFieldFig}(d).
If the distance $\rho$ to the EP vanishes, then $a_+\simeq a_-$ and $\phi_+\simeq \phi_-$ according to Eq.\,(\ref{SemiAxes}).
The ellipse excentricity vanishes ($\alpha\simeq\beta$) and the two mechanical eigenmodes coalesce into a single circularly polarized mode.
This is shown in Fig.\,\ref{MeanFieldFig}(e).
\section{Frequency modulation of the TLS drive}
\label{Appendix Frequency modulation of the TLS drive}

We start from the effective Hamiltonian of the driven TLS,
\begin{align}
\tilde H_{\rm TLS}(t) = 
\left( \begin{array}{cc}
\delta(t)/2 -g\,x(t) & \Omega_{\rm L}/2 \\
\Omega_{\rm L}/2 & -\delta(t)/2 + g\,x(t) \\
\end{array} \right) \,.
\end{align}
The luminescence excitation spectrum scales linearly with the excited-state population of the TLS.
The equations of motion for the reduced density matrix components $\sigma_{12}=\sigma_{21}^{*}$ and $\sigma_{22}=1-\sigma_{11}$ are obtained from $\tilde H_{\rm TLS}$ within the Born-Markov approximation.
They correspond to the usual optical Bloch equations,
\begin{align}
\left\{ \begin{array}{ll}
\dot \sigma_{12} &= -i[\delta(t) - 2g x(t) -i \Gamma/2]\, \sigma_{12} - i \Omega_{\rm L} (2\sigma_{22}-1)/2 \\
\dot \sigma_{22} &= \Omega_{\rm L} \Imag \sigma_{12} - \Gamma \sigma_{22}
\end{array} \right. \,.
\end{align}
They can be rewritten as
\begin{align}
\left\{ \begin{array}{ll}
\frac{dX}{d\tau}&=-i\left[D(\tau)-i/2\right]\,X -i\epsilon \left[ Y - 1/2 \right] \\
\frac{dY}{d\tau}&=-Y + \epsilon\Imag X
\end{array} \right. \,.
\end{align}
where we have introduced $X=\sigma_{12}=\sigma_{21}^{*}$, $Y=\sigma_{22}=1-\sigma_{11}$, $\tau = \Gamma t$, and $\epsilon=\Omega_{\rm L}/\Gamma$.
We then look for perturbative solutions of the type $X = \sum_{n} \epsilon^{n} X_{n}$ and  $Y = \sum_{n} \epsilon^{n} Y_{n}$ in the limit $\epsilon \ll 1$.
\subsection{Order n=0}
The Bloch equations lead to
\begin{align}
\left\{ \begin{array}{ll}
\frac{dX_{0}}{d\tau}&=-i\left[D(\tau)-i/2\right]\,X_{0} \\
\frac{dY_{0}}{d\tau}&=-Y_{0}
\end{array} \right. \,,
\end{align}
and the solutions generically read
\begin{align}
\left\{ \begin{array}{ll}
X_{0}&= X_{0}(\tau_{0})\, e^{-(\tau-\tau_{0})/2-i\int_{\tau_{0}}^{\tau}d\tau' D(\tau')} \notag \\
Y_{0}&= Y_{0}(\tau_{0})\, e^{-(\tau-\tau_{0})}
\end{array} \right. \,,
\end{align}
In the limit $\tau_{0}\rightarrow -\infty$, they reduce to
\begin{align}
\left\{ \begin{array}{ll}
\lim_{\tau_{0}\rightarrow -\infty} X_{0}&= 0 \\
\lim_{\tau_{0}\rightarrow -\infty} Y_{0}&= 0
\end{array} \right. \,.
\end{align}
\subsection{Order n=1}
The Bloch equations lead to
\begin{align}
\left\{ \begin{array}{ll}
\frac{dX_{1}}{d\tau}&=-i\left[D(\tau)-i/2\right]\,X_{1} -i \left[ Y_{0} - 1/2 \right] \\
\frac{dY_{1}}{d\tau}&=-Y_{1} + \Imag X_{0}
\end{array} \right. \,,
\end{align}
and the solutions generically read
\begin{align}
\left\{ \begin{array}{ll}
X_{1}&= X_{1}(0)\, e^{-i(\tau-\tau_{0})/2-i\int_{\tau_{0}}^{\tau}d\tau' D(\tau')} - i \int_{\tau_{0}}^{\tau}d\tau' \left[ Y_{0}(\tau') -1/2 \right]\, e^{(\tau'-\tau)/2+i\int_{\tau}^{\tau'}d\tau'' D(\tau'')} \\
Y_{1}&= Y_{1}(0)\, e^{-(\tau-\tau_{0})} + \int_{\tau_{0}}^{\tau} d\tau'  \Imag X_{0}(\tau')\, e^{(\tau'-\tau)}
\end{array} \right. \,.
\end{align}
In the limit $\tau_{0}\rightarrow -\infty$, they reduce to
\begin{align}
\left\{ \begin{array}{ll}
\lim_{\tau_{0}\rightarrow -\infty} X_{1}&= i/2 \int_{-\infty}^{\tau}d\tau'  e^{(\tau'-\tau)/2+i\int_{\tau}^{\tau'}d\tau'' D(\tau'')} \\
\lim_{\tau_{0}\rightarrow -\infty} Y_{1}&= 0
\end{array} \right. \,.
\end{align}
\subsection{Order n=2}
The Bloch equations lead to
\begin{align}
\left\{ \begin{array}{ll}
\frac{dX_{2}}{d\tau}&= - i \left[D(\tau)-i/2 \right]\,X_{2} - i \,Y_{1} \\
\frac{dY_{2}}{d\tau}&= - Y_ {2} + \Imag X_{1}
\end{array} \right. \,,
\end{align}
and the solutions generically read
\begin{align}
\left\{ \begin{array}{ll}
X_{2}&= X_{2}(0)\, e^{-(\tau-\tau_{0})2-i\int_{\tau_{0}}^{\tau}d\tau' D(\tau')} - i \int_{\tau_{0}}^{\tau}d\tau' Y_{1}(\tau')\, e^{(\tau'-\tau)/2+i\int_{\tau}^{\tau'}d\tau'' D(\tau'')} \\
Y_{2}&= Y_{2}(0)\, e^{-(\tau-\tau_{0})} + \int_{\tau_{0}}^{\tau} d\tau'  \Imag X_{1}(\tau')\, e^{(\tau'-\tau)}
\end{array} \right. \,.
\end{align}
In the limit $\tau_{0}\rightarrow -\infty$, they reduce to
\begin{align}
\left\{ \begin{array}{ll}
\lim_{\tau_{0}\rightarrow -\infty} X_{2}&= - i \int_{-\infty}^{\tau}d\tau' Y_{1}(\tau')\, e^{(\tau'-\tau)/2+i\int_{\tau}^{\tau'}d\tau'' D(\tau'')} \\
\lim_{\tau_{0}\rightarrow -\infty} Y_{2}&= \int_{-\infty}^{\tau} d\tau'  \Imag X_{1}(\tau')\, e^{(\tau'-\tau)}
\end{array} \right. \,.
\end{align}
At the end of the loop around the EP, either the flexural modes have gotten exchanged or they have not.
In both cases, the flexural dynamics is of the type $x(t)=Ae^{-\gamma_{i}t}\cos(\omega_{i}t+\phi_{i})$.
From now on, we consider $\phi_{i}=0$, which fixes an arbitrary origin of time.
The luminescence excitation spectrum scales linearly with
\begin{align}
\sigma_{22}(t) &\simeq \epsilon^{2} Y_{2}(t) \notag \\
&\simeq \frac{\Omega_{\rm L}^{2}}{2} e^{-\Gamma t} \Real \int_{-\infty}^{t}\!\!\!dt_{1}\, e^{(\frac{\Gamma}{2}-i\delta)t_{1}-i\beta_{\lambda}\sin(\omega_{\lambda}t_{1}+\phi_{\lambda})}
\int_{-\infty}^{t_{1}}\!\!\!dt_{2}\, e^{(\frac{\Gamma}{2}+i\delta)t_{2}+i\beta_{\lambda}\sin(\omega_{\lambda}t_{2}+\phi_{\lambda})}
e^{-i2g\int_{t_{1}}^{t_{2}}dt_{3}x(t_{3})} \,,
\end{align}
where $\beta_{\lambda}=a_{\lambda}/\omega_{\lambda}$, and we introduce the dimensionless parameter $\beta=2gA/\omega_{i}$.
In the limit $\beta\ll1$, the luminescence excitation spectrum can be approximated by
\begin{align}
\sigma_{22}(t)
&\simeq \frac{\Omega_{\rm L}^{2}}{4}\sum_{m,n}J_{m}(\beta_{\lambda})J_{n}(\beta_{\lambda})\Real [ e^{-i(m-n)\phi_{\lambda}} (A_{0}+A_{1}-A_{2}-A_{3}+A_{4})]
\end{align}
where the sum involves positive and negative values of the integers $m$ and $n$.
In addition, $J_{k}$ denotes the $k$th-order Bessel function of the first kind, and
\begin{align}
A_{0}&=\frac{e^{-i(m-n)\omega_{\lambda}t}}{(\delta+n\omega_{\lambda}-i\frac{\Gamma}{2})((m-n)\omega_{\lambda}+i\Gamma)} \notag \\
A_{1}&= \beta(1-i\frac{\gamma_{i}}{\omega_{i}}) \frac{e^{[-i(m-n)\omega_{\lambda}+i\omega_{i}-\gamma_{i}]t}}{(\delta+n\omega_{\lambda}-i\frac{\Gamma}{2})((m-n)\omega_{\lambda}-\omega_{i}+i\Gamma)} \notag \\
A_{2}&= \beta(1-i\frac{\gamma_{i}}{\omega_{i}}) \frac{e^{[-i(m-n)\omega_{\lambda}+i\omega_{i}-\gamma_{i}]t}}{(\delta+n\omega_{\lambda}+\omega_{i}-i\frac{\Gamma}{2})((m-n)\omega_{\lambda}-\omega_{i}+i\Gamma)} \notag \\
A_{3}&= \beta(1+i\frac{\gamma_{i}}{\omega_{i}}) \frac{e^{[-i(m-n)\omega_{\lambda}-i\omega_{i}-\gamma_{i}]t}}{(\delta+n\omega_{\lambda}-i\frac{\Gamma}{2})((m-n)\omega_{\lambda}+\omega_{i}+i\Gamma)} \notag \\
A_{4}&= \beta(1+i\frac{\gamma_{i}}{\omega_{i}}) \frac{e^{[-i(m-n)\omega_{\lambda}-i\omega_{i}-\gamma_{i}]t}}{(\delta+n\omega_{\lambda}-\omega_{i}-i\frac{\Gamma}{2})((m-n)\omega_{\lambda}+\omega_{i}+i\Gamma)} \,.
\end{align}
The term $A_{0}$ does not depend on $\beta$ and describes the luminescence excitation spectrum in the absence of electromechanical coupling.
We focus on $\Delta\sigma_{22}=\sigma_{22}\rfloor_{g\neq0}-\sigma_{22}\rfloor_{g=0}$.
We can then accumulate the luminescence over time, which results in
\begin{align}\label{Interferences}
\int_{0}^{\infty}dt\,\Delta\sigma_{22}(t)
&\simeq \beta\frac{\Omega_{\rm L}^{2}}{4}\sum_{m,n}J_{m}(\beta_{\lambda})J_{n}(\beta_{\lambda})\Real [ ie^{-i(m-n)\phi_{\lambda}} (B_{1}-B_{2}-B_{3}+B_{4})]
\end{align}
where
\begin{align}
B_{1}&= (1-i\frac{\gamma_{i}}{\omega_{i}})
\frac{\delta+n\omega_{\lambda}+i\Gamma/2}{(\delta+n\omega_{\lambda})^{2}+(\Gamma/2)^{2}}
\frac{(m-n)\omega_{\lambda}-\omega_{i}-i\Gamma}{((m-n)\omega_{\lambda}-\omega_{i})^{2}+\Gamma^{2}}
\frac{(m-n)\omega_{\lambda}-\omega_{i}+i\gamma_{i}}{((m-n)\omega_{\lambda}-\omega_{i})^{2}+\gamma_{i}^{2}} \notag \\
B_{2}&= (1-i\frac{\gamma_{i}}{\omega_{i}})
\frac{\delta+n\omega_{\lambda}+\omega_{i}+i\Gamma/2}{(\delta+n\omega_{\lambda}+\omega_{i})^{2}+(\Gamma/2)^{2}}
\frac{(m-n)\omega_{\lambda}-\omega_{i}-i\Gamma}{((m-n)\omega_{\lambda}-\omega_{i})^{2}+\Gamma^{2}}
\frac{(m-n)\omega_{\lambda}-\omega_{i}+i\gamma_{i}}{((m-n)\omega_{\lambda}-\omega_{i})^{2}+\gamma_{i}^{2}} \notag \\
B_{3}&= (1+i\frac{\gamma_{i}}{\omega_{i}})
\frac{\delta+n\omega_{\lambda}+i\Gamma/2}{(\delta+n\omega_{\lambda})^{2}+(\Gamma/2)^{2}}
\frac{(m-n)\omega_{\lambda}+\omega_{i}-i\Gamma}{((m-n)\omega_{\lambda}+\omega_{i})^{2}+\Gamma^{2}}
\frac{(m-n)\omega_{\lambda}+\omega_{i}+i\gamma_{i}}{((m-n)\omega_{\lambda}+\omega_{i})^{2}+\gamma_{i}^{2}} \notag \\
B_{4}&= (1+i\frac{\gamma_{i}}{\omega_{i}})
\frac{\delta+n\omega_{\lambda}-\omega_{i}+i\Gamma/2}{(\delta+n\omega_{\lambda}-\omega_{i})^{2}+(\Gamma/2)^{2}}
\frac{(m-n)\omega_{\lambda}+\omega_{i}-i\Gamma}{((m-n)\omega_{\lambda}+\omega_{i})^{2}+\Gamma^{2}}
\frac{(m-n)\omega_{\lambda}+\omega_{i}+i\gamma_{i}}{((m-n)\omega_{\lambda}+\omega_{i})^{2}+\gamma_{i}^{2}} \,.
\end{align}
Every term $B_{k}$ consists of a product of three Lorentzian functions.
The last one, which has the smallest width $\gamma_{i}(\ll\Gamma)$, results from the interferences due to the frequency modulation.
It describes narrow luminescence peaks of width $\gamma_{i}$ every time the condition $(m-n)\omega_{\lambda}=\omega_{i}$ is fulfilled.
For the typical values of the parameters that we are considering here, the nearly-degenerate flexural frequencies verify $|\omega_{1}-\omega_{2}|\ll \gamma_0$, so the two modes should be well resolved through the interference peaks in experiments.
\\
We can further estimate the characteristic amplitude of the interference peaks.
We consider that the laser frequency is modulated around the resonance ($\delta=0$), where the two flexural modes are not coupled.
The interference condition reads $(m-n)\omega_{\lambda}=\omega_{i}$ and requires $m\neq n$.
The main contribution in Eq.\,(\ref{Interferences}) involves the zeroth-order Bessel function, and $m$ and $n$ have to be as small as possible.
Thus, the main contributions when $\omega_{\lambda}=\omega_{i}$ arise from $B_{1}$ for $(m,n)=(1,0)$, $B_{2}$ for $(m,n)=(0,-1)$, $B_{3}$ for $(m,n)=(-1,0)$, and $B_{4}$ for $(m,n)=(0,1)$.
This leads to a peak of amplitude
\begin{align}
\int_{0}^{\infty}dt\,\Delta\sigma_{22}(t)
&\simeq - J_{0}(\beta_{\lambda})J_{1}(\beta_{\lambda})\left(\cos\phi_{\lambda}-\frac{\gamma_i}{\omega_{i}}\sin\phi_{\lambda}\right)\frac{\beta}{\gamma_{i}}\left(\frac{\Omega_{\rm L}}{\Gamma}\right)^{2}\,.
\end{align}
This explains the two interference peaks centered on the mechanical frequencies $\omega_1$ and $\omega_2$ in Fig.\,\ref{MonteCarlo}(b).

\twocolumngrid

\bibliographystyle{apsrev4-1}
\bibliography{references}

\end{document}